\documentclass
[prc,twocolumn,superscriptaddress,showpacs,twoside,floatfix]{revtex4}%
\usepackage{graphicx}
\usepackage{amsmath}

\begin{document}
\title{Two-proton radioactivity and three-body decay. IV. Connection to
quasiclassical formulation.}

\author{L.\ V.\ Grigorenko}
\affiliation{Flerov Laboratory of Nuclear Reactions, JINR, RU-141980 Dubna,
Russia}
\affiliation{Gesellschaft f\"{u}r Schwerionenforschung mbH, Planckstrasse 1,
D-64291, Darmstadt, Germany}
\affiliation{RRC ``The Kurchatov Institute'', Kurchatov sq.\ 1, 123182 Moscow,
Russia}

\author{M.\ V.\ Zhukov}
\affiliation{Fundamental Physics, Chalmers University of Technology, S-41296
G\"{o}teborg, Sweden}


\begin{abstract}
We derive quasiclassical expressions for the three-body decay width and
define the ``preexponential'' coefficients for them. The derivation is based on 
the integral formulae for the three-body width obtained in the  semianalytical
approach with simplified three-body Hamiltonian [L.V.\ Grigorenko and M.V.\
Zhukov, arXiv:0704.0920v1]. 
The model is applied to the decays of
the first excited $3/2^{-}$ state of $^{17}$Ne and $3/2^{-}$ ground state of
$^{45}$Fe. Various qualitative aspects of the model and relations with the other
simplified approaches to the three-body decays are discussed.
\end{abstract}


\maketitle


\section{Introduction}

\begin{figure}[ptb]
\includegraphics[width=0.48\textwidth]{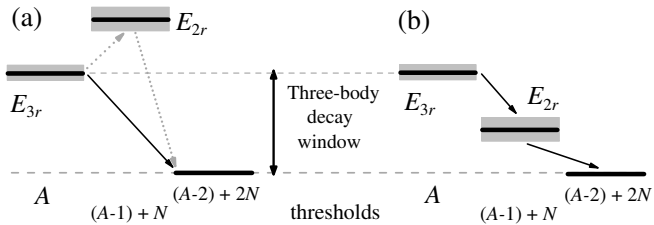}
\caption{Energy conditions for different modes of two-nucleon
emission (three-body decay): true three-body decay (a), sequential
decay (b).}
\label{fig:levels}
\end{figure}

The ``true'' two-proton decay \cite{gol60} is the decay mode which
is expected to be an ordinary phenomenon in the vicinity of the
proton dripline \cite{gri01a}. This mode corresponds to a specific
situation when one-proton emission is energetically (due to the
proton separation energy in the daughter system) prohibited and only
the simultaneous emission of two protons is possible. The energy
conditions are illustrated in  Fig.\ \ref{fig:levels}, more detailed
discussion of the three-body decay modes could be found in
\cite{gri01a,gri03c}. From theoretical point of view this situation
is a subset of the three-body Coulomb problem in the continuum. A
consistent quantum mechanical three-cluster model of the phenomenon
was developed in Refs.\ \cite{gri00b,gri01a,gri03c} and applied to a
range of the nuclear systems from $^{6}$Be to $^{66}$Kr in papers
\cite{gri02,gri02a,gri03,gri03a}. In the works \cite{gri05a,gri06}
possible importance of the ``true'' two-proton decay phenomenon was
demonstrated for astrophysical applications (in the form of the
reverse two-proton radiative capture process). Having in mind these
applications, which could require precise result in certain cases,
the semianalytical model with simplified three-body Hamiltonian was
developed in our recent work \cite{gri07}. The model allowed precise
calibrations of three-body calculations for decay widths.

On the other hand the semianalytical model \cite{gri07} can be used for 
derivation of the quasiclassical formulae for the three-body decay width. 
Formulae of this class can be found in early papers on two-proton radioactivity 
\cite{gol60,gal64}. They were used for qualitative estimates in our works 
\cite{gri01a,gri03,gri03a,gri03c,gol04,muk06}. We characterize them as 
quasiclassical as they are based on a certain factorization of the decay 
amplitude, which, in reality, is equivalent to existence of classical 
trajectories for system propagation in the process of the decay. A 
quasiclassical model for the width was introduced in the papers 
\cite{kry95,azh98} for $^{12}$O and then utilized in the series of works by 
Barker and Brown \cite{bar99,bar01,bro02,bar02,bar03,bro03} in somewhat modified 
form. It was characterized as ``R-matrix approach'', due to formal similarity 
with two-body R-matrix formalism.

\begin{figure}
\includegraphics[width=0.48\textwidth]{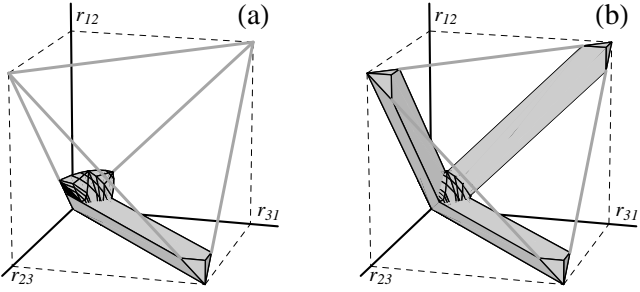}
\caption{Qualitative plots of the surface limiting the region of
classically allowed motion in the nuclear interior for three-body
problem. (a) One FSI and (b) three FSIs. Coordinates $r_{ij}$ are
pairwise distances for three constituents. The gray tetrahedron
limits the possible values of $r_{ij}$.}
\label{fig:shapes}
\end{figure}

The name of ``R-matrix approach'' can be, in certain sense, misleading as a 
derivation procedure is actually not based on the construction of the R-matrix 
on the ``nuclear surface''. The latter has a very complicated shape in the 
three-body system due to pairwise final state interactions (FSI). Fig.\ 
\ref{fig:shapes} shows schematically the surface limiting the region of 
classically allowed motion in the nuclear interior space for pairwise distances 
between clusters $r_{ij}$. Close to the origin this surface is approximated by 
the surface of constant hyperradius (ellipsoidal in this space). In the regions 
of final state interactions $r_{ij} \ll r_{ik},r_{kj}$ it is just 
$r_{ij}=\text{const}$. Possible values of radii due to triangle conditions 
$|r_{ik}-r_{kj}|\leq r_{ij} \leq r_{ik}+r_{kj}$ are limited by solid angle 
represented by gray tetrahedron in Fig.\ \ref{fig:shapes}. When particles 
penetrate through the Coulomb barrier they propagate predominantly along the 
``tunnels'' of classically allowed regions (though the motion entirely in the 
classically allowed region is not possible due to the energy condition of Fig.\ 
\ref{fig:levels}).

In this work we derive the quasiclassical expression for the three-body decay 
width in the case of the existence of narrow states in the subsystems and define 
the ``preexponential'' coefficients for the expression. The obtained expression 
has an important advantage compared to those used previously. We also make a 
comprehensive overview of quasiclassical three-body formulae used in different 
works and provide a qualitative analysis of them.

The unit system $\hbar=c=1$ is used in the article.

\section{Two-body case}


The integral formula for the decay width in two-body case is
\cite{kad71,gri07}
\begin{equation}
\Gamma=\frac{j_{l}}{N_{l}}=\frac{4}{v_{r}}\left\vert \int_{0}^{R} dr\;
\varphi_{l} (k_{r}r)\, \left(  \bar{V}-V \right)  \tilde{\psi}_{l}
(k_{r},r)\right\vert ^{2}\;,
\label{wid-2b}
\end{equation}
where the WF $\tilde{\psi}_{l}(k_{r},r)$ is the ``quasibound'' solution
for Hamiltonian
\[
(H-E_r)\tilde{\psi}_{l}=(T+V-E_r)\tilde{\psi}_{l}=0
\]
at the resonant energy $E_r$. ``Quasibound'' means that the solution
is matching the irregular (at the origin) Coulomb function $G_l$ in
the subbarrier region and it is normalized to unity in the finite
region of radius $R$. The function $\varphi_{l}(kr)$ is the
continuum WF of the auxiliary Hamiltonian $\bar{H}$
\[
(\bar{H}-E)\varphi_{l}=(T+\bar{V}-E)\varphi_{l}=0
\]
in the S-matrix representation having the asymptotic form
\begin{equation}
\varphi_{l}(kr)=\exp(i\bar{\delta}_l) \, \left[F_l(kr)
\cos(\bar{\delta}_l)+G_l(kr) \sin(\bar{\delta}_l)] \right] \;.
\label{wf-2b-as}
\end{equation}
The only restriction on the auxiliary Hamiltonian $\bar{H}$ is that its
eigenfunction should be sufficiently far from resonance at $E=E_r$.

The integral in Eq.\ (\ref{wid-2b}) can be rewritten in terms of Wronskian, see
Ref.\ \cite{gri07}:
\begin{eqnarray}
2M \!\!\int_{0}^{R}\!\!\varphi_{l}^{\ast}(V-\bar{V})\psi_{l}dr  =\left.  \left[
\varphi_{l}^{\ast}\!\left(  \frac{d}{dr}\psi_{l}\right)  -\left(  \frac{d}
{dr}\varphi_{l}^{\ast}\right) \! \psi_{l}\right] \right\vert _{r=R} \nonumber \\
\qquad =-i\exp(-i\bar{\delta}_{l})\;\cos(\bar{\delta}_{l}) \;
k_{r}W(F_{l},G_{l}) \;.
\quad \nonumber
\end{eqnarray}
The function $\psi_{l}$ on the left from $R$ is obtained as a
solution of the Schr\"odinger equation [normalized only by
asymptotic condition $\Psi_l=G_{l}$]. To the right of the
matching point it can be written explicitly in the form correctly
normalized both asymptotically (to $G_{l}$) and in the internal
region (to unity):
\[
\tilde{\psi}_{l}(k_{r}r)\overset{r>R}{=}\left(  \int_{0}^{R} \left \vert
\psi_{l}(k_{r}r)\right \vert ^{2}dr \right )  ^{-1/2}\frac{\psi_{l}(k_{r}
R)}{G_{l}(k_{r}R)}G_{l}(k_{r}r)\;.
\]
Neglecting the terms of the order $\bar{\delta}_{l}$ we obtain
\[
\Gamma=\frac{1}{MR^{2}}\frac{\left \vert \psi_{l}(k_{r}R)\right \vert ^{2}}
{\int_{0}^{1}\left \vert \psi_{l}(k_{r}Rx)\right \vert ^{2}dx}\frac{k_{r}R}
{G_{l}^{2} (k_{r}R)}\;.
\]
For small widths the relation $G_{l}\gg F_{l}$ is true and the
penetrability can be identified as
\[
P_{l}(k_{r},R)=\frac{k_{r}R}{F_{l}^{2} (k_{r}R)+G_{l}^{2}(k_{r}R)}\approx
\frac{k_{r}R}{G_{l}^{2}(k_{r}R)}\;.
\]
Then we obtain for the width in Eq.\ (\ref{wid-2b}) the standard
R-matrix expression
\begin{equation}
\Gamma=2\gamma^{2}P_{l}(k_{r},R)=2\gamma_{WL}^{2}\theta^{2}P_{l}
(k_{r},R)\;,
\label{r-mat-wid-2}
\end{equation}
where the Wigner limit for the reduced width $\gamma_{WL}^{2}$ and the
dimensionless reduced width (DRW) $\theta^{2}$ are identified as
\begin{equation}
\gamma_{WL}^{2}=\frac{1}{2MR^{2}}\;,\quad \theta^{2}=\frac{\left \vert \psi_{l}
(k_{r}R) \right \vert ^{2}}{\int_{0}^{1}\left \vert \psi_{l}(k_{r}xR) \right
\vert ^{2}dx}\;.
\label{dimles-red-wid}
\end{equation}

The DRW is expected to be of the order of unity for $R$ chosen in
the subbarrier region. However, the DRW $\theta^{2}$ decreases
relatively strongly with radius $R$  for two reasons: (i) the
normalization integral in the denominator is growing with $R$
\footnote{This growth is significant for $R$ chosen in a relatively
broad subbarrier range (which depends certainly on the barrier
strength). For $R$ formally chosen beyond the barrier there will be
slow, more or less linear growth of the normalization with $R$ which
we should regard as unphysical.}; (ii) the numerator of Eq.\
(\ref{dimles-red-wid}) is decreasing as $G_l^{2}$ with $R$ under the
barrier. This decrease is compensated in Eq.\ \ref{r-mat-wid-2} by
growth of the penetrability as $1/G_l^{2}$. Thus for values of $R$
under the barrier the width  should decrease slowly with radius $R$
up to some value of the radius (close to the outer classical turning
point of the barrier), where it is practically stabilized.


\subsection{Integration region in calculations of widths}


The role of the integration limit in the calculations of the
internal normalization can be understood using the R-matrix example.
Similar calculations can be found in the book \cite{obormot}, where
they illustrate a somewhat different issue. In the case of a square
well potential and absence of Coulomb interaction the normalization
integral for the internal region is easily evaluated
\begin{eqnarray}
N_{1} &= & \int_{0}^{1}\left[  (kRx)j_{l}(kRx)\right]  ^{2}dx=\frac{(kR)^{2}}
{2}[j_{l}(kR)^{2}  \nonumber \\
& - & j_{l+1}(kR)j_{l-1}(kR)]\approx\frac{(k_{r}R)^{2}}{2}\left[  j_{l}
(k_{r}R)\right]  ^{2}\;.
\label{norm-teta-in}
\end{eqnarray}
The approximate equality is good at resonance and becomes exact if
$E_{r}\rightarrow0$. The dimensionless reduced width is in that case
\[
\theta_{1}^{2}=2\left(  1-\frac{j_{l+1}(kR)j_{l-1}(kR)}{\left[  j_{l}
(kR)\right]  ^{2}}\right)  ^{-1}\;.
\]
We can redefine the DRW in such a way that penetrability could be
evaluated at the nuclear surface but the effect of the contribution
to normalization from the subbarrier region is already taken into
account. The normalization integral is reasonable to calculate up to
the first zero of the irregular function $G_{l}(k_{r}r)$. The WF
comes from under the barrier approximately at this point and
calculation of the normalization integral beyond this point loose a
sense (the particle can not anymore be considered to be in the
``internal region'').

Beyond the matching point $R$ the function can be approximated as
\[
G_{l}(k_{r}r)=(k_{r}r)n_{l}(k_{r}r)\sim r^{-l}\;,
\]
with a good precision and integration in Eq.\ (\ref{dimles-red-wid}) can be
extended to infinity. The integral of such approximated function converges for
integration up to infinity for $l>0$. Thus the normalization for the whole
subbarrier domain is
\begin{eqnarray}
N_{2}& = & N_{1}+\left[  (kR)j_{l}(kR)\right]  ^{2}\int_{1}^{\infty} x^{-2l}dx
\nonumber  \\
&  \approx & \frac{2l+1}{2l-1}\frac{(k_{r}R)^{2}}{2}\left[  j_{l}(k_{r}R)\right]
^{2}\;.
\label{norm-teta-out}
\end{eqnarray}
The ``redefined'' dimensionless reduced width is exactly written as
\[
\theta_{2}^{2}=2\left(  \frac{2l+1}{2l-1}-\frac{j_{l+1}(kR)j_{l-1}
(kR)}{\left[  j_{l}(kR)\right]  ^{2}}\right)  ^{-1}\;.
\]
The ratio of the reduced widths calculated at square well boundary by Eq.\
(\ref{norm-teta-in}) and for `` effective infinity'' by Eq.\
(\ref{norm-teta-out}) is approximately
\[
\theta_{2}^{2}/\theta_{1}^{2} \approx (2l-1)/(2l+1)\;,
\]
again the expression becomes an identity for $E_r \rightarrow 0$.

In Table \ref{tab:r-mat-2} the mentioned values are provided for the
simple model employed here. We can find out that (i) the effect of
the subbarrier region is \emph{very} important for weak barriers
(e.g.\ $l=1$ in our example) and it gradually diminish as the
barrier grows; (ii) the effect is not absolutely negligible even for
quite high barriers; (iii) the direct numerical comparison  shows
that the DRW value $\theta_{2}^{2}$ calculated for ``effective
infinity'' is exactly consistent with the width obtained in
scattering calculations, and should be considered as a correct
definition.


\subsection{Evaluation of the integral in Eq.\ (\ref{wid-2b})}
\label{sec:r-mat-two-body}


If the width of the resonance in the auxiliary Hamiltonian is sufficiently
small, then, in proximity of the auxiliary resonance energy $E_{a}$, we can
write confidently that in the internal region
\begin{equation}
\varphi_{l}(kr)=\sqrt{\frac{\pi v}{2}}\sqrt{D_{l}\,\frac{\Gamma_{a}(E)/2\pi}
{(E-E_{a})^{2}+\Gamma_{a}^{2}(E)/4}}\;\hat{\psi}_{l}(k_{a},r)\;,
\label{res-fun}
\end{equation}
where function $\hat{\psi}_{l}(k_{a},r)$ is also the quasibound WF
as $\tilde{\psi}_{l}(k_{r},r)$ (namely with resonant boundary
condition and normalized to unity in the internal domain) but taken
at the resonant energy $E_a$ of the auxiliary Hamiltonian
\[
\hat{\psi}_{l}(k_{a},r>R)\propto G_{l}(k_{a}r)\;,\quad\int_{0}^{R}
dr\;\left\vert \hat{\psi}_{l}(k_{a},r)\right\vert ^{2}=1\;.
\]
The coefficient $D_{l}$ provides the normalization of the Breit-Wigner profile,
which is already practically normalized
\[
\int_{-\infty}^{\infty}dE\frac{\Gamma_{a}(E_{a})/2\pi}{(E-E_{a})^{2}%
+\Gamma_{a}^{2}(E_{a})/4}=1\;,
\]
within the energy domain of interest:
\[
1=D_{l}\int_{0}^{E_{a}+\Delta E} dE \frac{\Gamma_{a}(E)/2\pi}{(E-E_{a})^{2} +
\Gamma_{a} (E)^{2}/4}\;.
\]
For reasonably narrow states $D_{l}\equiv 1$ means that $\Delta
E=(3-5)\Gamma _{a}(E).$ So this coefficient is always reasonably
close to 1. The coefficient $\sqrt{\pi v/2}$ in the Eq.\
(\ref{res-fun}) is partly connected with normalization of the
partial radial functions $\varphi_{l}(kr)$
\[
\int_{0}^{\infty}\varphi_{l}^{\ast}(k^{\prime}r)\varphi_{l}(kr)dr=\frac{\pi
}{2}\delta(k^{\prime}-k)\;,
\]
and partly with conversion from integration over $dk$ to integration
over $dE$. Now substituting (\ref{res-fun}) into (\ref{wid-2b}) and
using (\ref{r-mat-wid-2}) the identity
\begin{equation}
\Gamma(E_{r})=\Gamma(E_{r})\;\frac{\theta_{a}^{2}D_{l}}{\theta^{2}}
\frac{ \left \vert \int_{0}^{R} dr \;\hat{\psi}_{l}^{\ast}(k_{a},r)\left(
\bar{V}-V \right)  \tilde{\psi}_{l}(k_{r},r) \right \vert ^{2}}{(E_{r}-E_{a}
)^{2}+\Gamma_{a}(E_{r})^{2}/4}.
\label{width-identity}
\end{equation}
is obtained.

If the difference between energies $E_{r}-E_{a}$ is small compared
to the height of the barrier, the quasibound WFs $\hat{\psi}_{l}$
and $\tilde{\psi }_{l}$ should be quite close to each other and
provide close DRW values
\begin{equation}
E_{r}-E_{a}\ll E_{bar}\; , \; \hat{\psi}_{l}(k_{a},r)\approx\tilde{\psi}
_{l}(k_{r},r)\; \rightarrow \;\theta_{a}^{2}/\theta^{2}\approx 1 \,.
\label{rmat-ass}
\end{equation}
Also the variation of the kinetic energy should be small in such a
case and the whole variation in energy should be due to the
potential energy change and it should be true that
\begin{equation}
\frac{\theta_{a}^{2}}{\theta^{2}}\left\vert \int_{0}^{R} dr\; \hat{\psi}_{l}
^{\ast} (k_{a},r) \left(  \bar{V}-V\right)  \tilde{\psi}_{l}(k_{r},r) \right
\vert^{2} = (E_{r}-E_{a})^{2}B_{l}\;,
\label{en-dep-2}
\end{equation}
with coefficient $B_{l} \approx 1$. The examples of actual
calculated values of these coefficients are provided in Table
\ref{tab:r-mat-2} for the model of the previous Section. They
indicate that the auxiliary resonance $E_{a}$ should be sufficiently
narrow and sufficiently close to the physical resonance $E_{r}$ to
make the approximation of Eq.\ (\ref{res-fun}) really precise. In
the opposite case the value $D_{l}$ needs to be correspondingly
renormalised to preserve the identity (\ref{width-identity}). Anyhow
we keep in mind that for states broader than $10^{-2}$ MeV the
substitution (\ref{res-fun}) provides results which are valid only
within a factor of 2.

The issues discussed above are not of importance for the standard
R-matrix phenomenology, but they should be clearly understood before
we turn to the three-body case.


\section{Three-body case}
\label{sec:three-body-rmat}


In paper \cite{gri07} we used the simplified three-body Hamiltonian
in which the proton-proton FSI was neglected \footnote{It was shown
in \cite{gri07} that the assumption of the ``diproton decay''
provides a negligible width compared to the assumption of the
``sequential decay''. This means that it is very natural to neglect
$p$-$p$ interaction while constructing the simplified model of the
decay.}. In this Hamiltonian we used also a simplified expression
for the three-body Coulomb interaction, which allows an isolation of
the degrees of freedom. The three-body width in this simplified
model was obtained as
\begin{equation}
\Gamma(E_{3r})=\frac{8E_{3r}}{\pi}\int\nolimits_{0}^{1}d\varepsilon\;
\frac{M_xM_y} {k_{x}(\varepsilon) k_{y}(\varepsilon)} \left\vert
A(\varepsilon)\right\vert ^{2} \;,
\label{width-3b}
\end{equation}
where energy and momenta of the subsystems are
\[
E_x=\varepsilon E_{3r}\; , \quad
E_y=(1-\varepsilon) E_{3r}\; , \quad
k_i=\sqrt{2M_i E_i}\; .
\]
The ``energy distribution'' coefficient is defined
\begin{eqnarray}
A(\varepsilon) &  = & \int\nolimits_{0}^{\infty}dXdY\;\varphi_{l_{x}}
(k_{x}(\varepsilon)X)\;\varphi_{l_{y}}(k_{y}(\varepsilon)Y)
\nonumber \\
&  \times & V_{3}(X,Y)\;\tilde{\varphi}_{Ll_{x}l_{y}S}(k_{r},X,Y)\;,
\label{coef-a}
\end{eqnarray}
where the function $\tilde{\varphi}_{Ll_{x}l_{y}S}(k_{r},X,Y)$ is
the radial part of the solution with the outgoing asymptotic
behaviour for a simplified three-body Hamiltonian
\begin{eqnarray}
\left(H  -  \tilde{E}_{3r}\right)\tilde{\varphi} = \left(H_x+H_y
+V_{3}(\rho)-\tilde{E}_{3r}\right) \tilde{\varphi} \qquad \nonumber \\
\quad =  \left(T_x+T_y+V_x(X)+V_y(Y)+V_{3}(\rho)-\tilde{E}_{3r} \right)
\tilde{\varphi}=0 \nonumber
\end{eqnarray}
at the complex pole energy $\tilde{E}_{3r}=E_{3r}+i\Gamma/2$. The
three-body potential $V_{3}(\rho)$, which is depending only on the
hyperradius $\rho$, is used to form the three-body resonance and
control the resonant energy.  The functions $\varphi_{l_{x}},
\varphi_{l_{y}}$ are the continuum eigenfunctions of the separable
auxiliary Hamiltonian $\bar{H}=H_x+H_y$ and they
are normalized by the asymptotic condition
(\ref{wf-2b-as}) where phase shifts are defined for the
subhamiltonians
\begin{eqnarray}
(T_x+V_x(X)-E_x)\varphi_{l_x}(k_xX) & = & 0 \; , \nonumber \\
(T_y+V_y(X)-E_y)\varphi_{l_y}(k_yY) & = & 0 \; . \nonumber
\end{eqnarray}

In the case when there are narrow resonant states in both ``X'' (at
energy $E_{xa}$) and ``Y'' (at energy $E_{ya}$) subsystems, the
substitution (\ref{res-fun}) can be used for both $\varphi_{l_{x}}$
and $\varphi_{l_{y}}$. It is clear that this approximation is
physical only when the so-called ``Y'' Jacobi system is chosen. In
such a Jacobi system $X$ is a coordinate between the core and a
proton, $Y$ is a distance between the ``X'' subsystem center of mass
and the second proton. The hyperradius is then defined as
\[
\rho^2=\frac{A_c}{A_c+1}\;X^2 + \frac{A_c+1}{A_c+2}\;Y^2 \; ,
\]
where $A_c$ is the core mass. This approximation is good enough only
in relatively heavy nuclei where the $Y$ Jacobi coordinate is close
to the single-particle proton coordinate due to a large core mass.
The total auxiliary Hamiltonian $\bar{H}$ has a resonant energy
\[
E_{a}=E_{xa}+E_{ya}\;,\qquad k_{ia}=\sqrt{2M_iE_{ia}}\;.
\]
Denoting the following integral as
\begin{eqnarray}
\left\langle V_{3}\right\rangle  & = & \int_{0}^{\infty}dXdY\;\hat{\varphi
}_{l_{x}}^{\ast}(k_{xa},X)\,\hat{\varphi}_{l_{y}}^{\ast}(k_{ya},Y)\ V_{3}
(\rho)\;
\nonumber \\
& \times & \tilde{\varphi}_{Ll_{x}l_{y}S}(k_{r},X,Y)\;,
\nonumber
\end{eqnarray}
where $\hat{\varphi}$ are normalized quasibound WFs for the subsystems, the
width is obtained as
\begin{eqnarray}
\Gamma(E_{3r}) &  = & \frac{E_{3r}\left \langle V_{3}\right \rangle ^{2}} {2\pi}
\int \nolimits_{0}^{1}d \varepsilon \; \frac{D_{x}\; \Gamma_{xa}(E_{x})} 
{(E_{x}-E_{xa})^{2}+\Gamma_{xa}(E_{x})^{2}/4}
\nonumber \\
&  \times & \frac{D_{y}\; \Gamma_{ya}(E_{y})} {(E_{y}-E_{ya})^{2} + \Gamma_{ya} 
(E_{y})^{2}/4}\;.
\label{rmat-wid-3}
\end{eqnarray}
The expression for the two-proton width was, for the first time,
obtained in a similar form in Ref.\ \cite{gal64}. It is difficult to
say, why this work did not attract attention and why its results
have never been used. A possible reason could be the relatively
complicated procedure used in Ref.\ \cite{gal64} and a lack of
qualitative investigation of the model properties and the nature of
approximations involved.

Using the following notations
\begin{eqnarray}
\Gamma_{i}(E_{i})=2\gamma_{i}^{2}P_{l_{i}}(E_{i},r_{chi},Z_{i})=\frac
{\theta_{i}^{2}}{M_{i}r_{chi}^{2}}\,P_{l_{i}}(E_{i},r_{chi},Z_{i})\;,
\nonumber \\
\mathcal{P}_{l_{x}l_{y}}(\varepsilon,E)=P_{l_{x}}(\varepsilon
E,r_{chx},Z_{x})\;P_{l_{y} }((1-\varepsilon)E,r_{chy},Z_{y})\;,
\nonumber \\
\Delta_{l_{i}}(\varepsilon,E,E_{ia})=\left[  (1-\frac{\varepsilon E}
{E_{ia}})^{2} + \frac{\Gamma _{ia}(\varepsilon E)^{2}}{4E_{ia}^{2}}\right]
^{-1}\;, \nonumber
\end{eqnarray}
we can rewrite Eq.\ (\ref{rmat-wid-3}) in the ``dimensionless'' form
for the penetration part of the expression: 
\begin{eqnarray}
\Gamma(E_{3r}) &  = & \frac{D_{x}D_{y}}{2\pi} \frac{2 \gamma_{x}^{2}2 
\gamma_{y}^{2} E_{3r}} {E_{xa}^{2}E_{ya}^{2}}\,\left \langle V_{3} \right 
\rangle ^{2} \int \nolimits_{0}^{1}d \varepsilon \,\mathcal{P}_{l_{x}l_{y}} 
(\varepsilon,E_{3r})
\nonumber \\
&  \times & \Delta_{l_{x}} (\varepsilon,E_{3r},E_{xa}) \,\Delta_{l_{y}}
(1-\varepsilon,E_{3r},E_{ya}) \;.
\label{rmat-wid-dms-3}
\end{eqnarray}
The expression for width is now explicitly factorized into ``preexponent'' 
(which has the dimension of energy) and dimensionless ``exponential'' part
\[
\int \nolimits_{0}^{1} d \varepsilon \,\mathcal{P}_{l_{x}l_{y}}\Delta_{l_{x}} 
\Delta_{l_{y}} \sim \exp \left[ - \sqrt{M/E_{3r}} \, 2 Z_{\text{core}} C \right] 
\;, 
\]
where $C$ is coefficient of the order of unity. 

Following the discussion of Eq.\ (\ref{en-dep-2}) we can write
\begin{equation}
\left\langle V_{3}\right\rangle ^{2}=(E_{3r}-E_{xa}-E_{ya})^{2}\; D_{3}\;,
\label{v3-en-dep}
\end{equation}
where the dimensionless coefficient $D_{3}$ is presumed to be close
to unity. Finally we get for the width:
\begin{eqnarray}
\Gamma(E_{3r})=\frac{D_{x}D_{y}D_{3}\theta_{x}^{2}\theta_{y}^{2}}{2\pi}
\frac{E_{3r}(E_{3r}-E_{xa}-E_{ya})^{2}} {M_{x}M_{y} r_{chx}^{2}r_{chy}^{2}
E_{xa}^{2} E_{ya}^{2}} \int\nolimits_{0}^{1}d\varepsilon \nonumber \\
\times  \mathcal{P}_{l_{x}l_{y}} (\varepsilon,E_{3r})\Delta_{l_{x}
}(\varepsilon,E_{3r},E_{xa}) \Delta_{l_{y}}(1-\varepsilon,E_{3r},E_{ya}) \quad
\label{rmat-wid-dms-3s}
\end{eqnarray}

The $D_{3}$ values calculated in the simplified three-body model are given in 
Table \ref{tab:wid-three-d}. It can be shown that the ratio $\left\langle 
V_{3}^{2} \right \rangle /\left\langle V_{3}\right\rangle ^{2}$ is a measure of 
the WF to be outside the interaction region (this statement is trivial to check 
for the square well potential). The values of $\left\langle V_{3}^{2} \right 
\rangle $ and $\left\langle V_{3}\right\rangle ^{2}$ calculated in the 
three-body model for $^{17}$Ne and $^{45}$Fe are also provided in Table 
\ref{tab:wid-three-d}. The ratios $\left\langle V_{3}^{2}\right\rangle 
/\left\langle V_{3}\right\rangle ^{2}$ are reasonably consistent with the 
$D_{3}$ values and are quite close to unity. This indicates that the WFs are 
predominantly localized in the ``internal'' regions.


\section{Discussion}


Table \ref{tab:wid-sens} demonstrates a sensitivity of three-body widths 
estimated by quasiclassical expressions to different ingredients of the models. 
However, before discussing these effects we should make some overview of the 
models.


\subsection{Our previous quasiclassical model}


In the pioneering work of Goldansky \cite{gol60} the differential probability of 
the two-proton decay was estimated as
\[
w(E_{3r},\varepsilon)\sim\exp\left[  -\frac{2\pi\,Z_{\text{core}}\sqrt{M}}
{\sqrt{E_{3r}}}\left(  \frac{1}{\sqrt{\varepsilon}}+\frac{1}
{\sqrt{1-\varepsilon}}\right)  \right]  \;.
\]

In our works \cite{gri01a,gri02,gri03,gri03a,gri03c} the quasiclassical
expression for the two-proton width was used in the form
\begin{equation}
\Gamma_{s}(E_{3r})=\frac{6E_{3r}^{1/2}}{\pi(r_{chx}r_{chy})^{3/2}(M_{y}M_{x})^{
3/4} } \int\nolimits_{0}^{1} d \varepsilon \mathcal{P}_{l_{x}l_{y}}
(\varepsilon,E_{3r})\,.
\label{wid-3-prev}
\end{equation}
The ``exponential'' coefficient
$\mathcal{P}_{l_{x}l_{y}}(\varepsilon,E_{3r})$ is closely related to
the function $w(E_{3r},\varepsilon)$ above. The motivation for the
preexponent choice was like follows. Let's consider the two-body
case in the situation of no barrier (no Coulomb interaction and zero
angular momentum $l=0$):
\[
\Gamma=2 \gamma ^{2}P_{l}(k,r_{ch},Z) = 2 \frac{\theta^{2}(kr_{ch})}
{2Mr_{ch}^{2}} \overset{\theta^{2} \rightarrow2} {=} \frac{2v}{r_{ch}} =
\frac{1}{\tau (r_{ch}/2)}\;.
\]
The width in that case is formally just the inverse flight time for distance
$r_{ch}/2$ [denoted as $\tau(r_{ch}/2)$ above]. Using the same assumption (no
Coulomb and zero angular momentuma $l_x=l_y=0$) we obtain from Eq.\
(\ref{wid-3-prev})
\[
\Gamma_{s}(E_{3r})  =  \frac{6r_{chx}r_{chy}E_{3r}^{1/2}} { \pi (r_{chx}
r_{chy})^{3/2} (M_{y} M_{x})^{3/4}} \int \nolimits_{0}^{1} d \varepsilon \;
k_{x}k_{y} \;.
\]
Using the integral value
\[
\int\nolimits_{0}^{1} d \varepsilon
\;\sqrt{\varepsilon(1-\varepsilon)}=\frac{\pi}{8}\;,
\]
the estimate for the width is obtained as
\[
\Gamma_{s}(E_{3r})=\frac{3E_{3r}^{1/2}}{2(r_{chx}r_{chy})^{1/2}
(M_{y}M_{x})^{1/4}} \approx \frac{3}{\sqrt{8}}\frac{v}{r_{ch}} \approx \frac{1}{
\tau (r_{ch})}\;.
\]
Thus, the width in such a ``no barrier'' case is normalized to the
inverse flight time to some ``typical internal distance'' $r_{ch}$.

The above idea of the preexponent derivation can be found too
simplistic. However, calculations show that for the considered cases
Eq.\ (\ref{wid-3-prev}) demonstrates a good consistency with Eq.\
(\ref{rmat-wid-dms-3s}) when the channel radii are chosen in a sound
way (namely they should be close to the inner classical turning
point of the Coulomb barrier \cite{turning}).


\subsection{Special cases of the present model}


Let's consider the some special cases of Eq.\ (\ref{rmat-wid-dms-3s}).

\subsubsection{True three-body decay, $E_{xa}\gg E_{3r}$, $E_{ya}\gg E_{3r}$}

In that case
the main contribution to the energy integral in (\ref{rmat-wid-dms-3s}) is 
connected with $\varepsilon= 1 /2$. Replacing the slowly varying functions 
$\Delta$ by their constant values at $\varepsilon= 1 /2$ we obtain
\begin{eqnarray}
\Gamma_{m}(E_{3r}) = \frac{E_{3r}(E_{3r}-E_{xa} -E_{ya})^{2}}
{(E_{xa}-E_{3r}/2)^{2} (E_{ya}-E_{3r}/2)^{2}} \qquad \nonumber \\
\times \; \frac{2}{\pi} \, D_{x} D_{y} D_{3} \, \gamma_{x}^{2} \gamma_{y}^{2}
\; \int \nolimits_{0}^{1} d \varepsilon \; \mathcal{P}_{l_{x}l_{y}}
(\varepsilon,E_{3r})\;. \quad
\label{gamma-3-m}
\end{eqnarray}
Thus we obtain formula analogous to (\ref{wid-3-prev}), used in our
previous works, but with different preexponent. It should be noted:

\noindent (i) The preexponent in this form is explicitly depending on the
resonance properties of the subsystems.

\noindent (ii) It can be seen in Table \ref{tab:wid-sens} that
agreement is very good between the value $\Gamma$ in Eq.\
(\ref{rmat-wid-dms-3s}) and the approximation $\Gamma_{m}$ in Eq.\
(\ref{gamma-3-m}), calculated neglecting the variation of functions
$\Delta_i$ within the decay window.

\noindent (iii) The dimensionless coefficient $D_{x}D_{y}D_{3}$ is
not that different from unity (at the level of $10-50 \%$).
The  $D_{x}$ and $D_{y}$ coefficients should be very close to unity
for sufficiently narrow states, so we need to know only the
properties of the two-body subsystem to fix this ingredient of the
model. However, the $D_{3}$ coefficient is beyond the R-matrix
ideology and requires a validation within the three-body model.

\noindent (iv) If both subsystems (on $X$ and $Y$ coordinates) are equivalent,
then $E_{xa} \equiv E_{ya}$ and the three-body width dependence on the two-body
resonance position should have the following typical dependence:
\begin{equation}
\Gamma(E_{3r})\sim (E_{xa}-E_{3r}/2)^{-2} \;.
\label{ass-rel-wid}
\end{equation}
This dependence we have observed in the calculations within the simplified
three-body model for $^{45}$Fe (two equivalent final state interactions), see
Fig.\ 14 of Ref.\ \cite{gri07}.


\subsubsection{Sequential decay, $E_{xa}<E_{3r}$, $E_{ya}\gg E_{3r}$}


It is quite simple to integrate over $d\varepsilon$ in the Eq.\
(\ref{rmat-wid-3})  for these energy conditions:
\begin{eqnarray}
\Gamma(E_{3r}) & \approx & \frac{D_{y}\;\left \langle V_{3}\right \rangle ^{2}}
{(E_{3r}-E_{xa}-E_{ya})^{2}}\;\Gamma_{ya}(E_{3r}-E_{xa})
\nonumber \\
\Gamma(E_{3r}) & \approx &  D_{y}D_{3}\;\Gamma_{ya}(E_{3r}-E_{xa}) \nonumber
\end{eqnarray}
Thus the width is reduced to a two-body expression with some
modifications, which takes into account the three-body character of
the model considered for the resonant state.


\subsubsection{True three-body decay, $E_{xa}>E_{3r}$, $E_{ya}\gg E_{3r}$}


In this case we can approximate the $(E_{3r}-E_{xa}-E_{ya})^{2}$ by $E_{ya}^{2}$ 
in the numerator and $(E_{ya}-E_{3r}/2)^{2}$ by $E_{ya}^{2}$ in the denominator 
and obtain
\begin{equation}
\Gamma(E_{3r})=\frac{2}{\pi}\,\frac{D_{x}D_{y}D_{3} \; \gamma_{x}^{2}\,
\gamma_{y}^{2}\;E_{3r}}{(E_{xa}-E_{3r}/2)^{2}} \!
\int \nolimits_{0} ^{1} \! \! \varepsilon \,\mathcal{P}_{l_{x}l_{y}}
(\varepsilon,E_{3r}) \;.
\label{wid-ofsi}
\end{equation}
This situation is close to the one final state interaction (OFSI)
model considered in Ref.\ \cite{gri07} for methodological purposes.
The three-body width dependence on the two-body resonance position
is again
\[
\Gamma(E_{3r})\sim (E_{xa}-E_{3r}/2)^{-2} \;.
\]
as in Eq.\ (\ref{ass-rel-wid}) where two equivalent FSIs are
considered. This behaviour is reasonably well reproduced for
$^{17}$Ne within the simplified three-body OFSI model, see Fig.\ 5
in Ref.\ \cite{gri07}.

It should, however, be kept in mind that the considered
approximation is valid and precise if the energy of the higher
resonance $E_{ya}$ is kept significantly below the barrier. It is
necessary for two reasons. (i) Geometric proximity of the resonance
WFs $\hat{\psi}_{l_{x}l_{y}}^{\ast}(X,Y)$ and
$\tilde{\psi}_{l_{x}l_{y}}(X,Y)$ is requred for Eq.\
(\ref{v3-en-dep}) which is based on the smallness of kinetic energy
contribution to the variation of total energy. (ii) The width of the
upper resonance should be sufficiently small as demonstrated in
Section \ref{sec:r-mat-two-body}. All the mentioned conditions for
this special case are difficult to meet in practice. It can be seen
in Fig.\ 5 of Ref.\ \cite{gri07} that the width calculated for the $^{17}$Ne
$3/2^-$ state in OFSI model (one subsystem is nonresonant) follows
the analytical dependence Eq.\ (\ref{ass-rel-wid}) with significant
deviations, while in the case of calculations for $^{45}$Fe shown in
Fig.\ 13 \cite{gri07} (both subsystems have resonances) the
agreement is practically perfect.

A quasiclassical expression for the width was introduced in the
papers \cite{kry95,azh98} for $^{12}$O in the form
\begin{equation}
\Gamma_{b}(E_{3r})=\frac{2D_{x}}{\pi}\frac{\gamma_{x}^{2}\gamma_{y}^{2} E_{3r}}
{E_{xa}^{2}} \! \int \nolimits_{0}^{1}  \! d \varepsilon
\mathcal{P}_{l_{x}l_{y}}(\varepsilon,E) \Delta _{l_{x}}
(\varepsilon,E_{3r},E_{xa}) \;,
\label{wid-3-bark}
\end{equation}
(given here in our notations), which is practically equivalent to
Eq.\ (\ref{wid-ofsi}). It was introduced without derivation and
analysis of the involved approximations. The expression was
utilized in the series of works
\cite{bar99,bar01,bro02,bar02,bar03,bro03} by Barker and Brown in
somewhat modified form. We can see in Table \ref{tab:wid-sens} that
for the systems under consideration the disagreement between
``correct'' value $\Gamma$ Eq.\ (\ref{rmat-wid-dms-3s}) and
``special case'' $\Gamma_{b}$ Eq.\ (\ref{wid-3-bark}) is as large as
a factor $2-4$. It is also easy to show analytically that for
systems with $E_{xa} \equiv E_{ya}$ (that means for all the ground
state decays) the difference should be about a factor 4.


\subsection{Diproton model}


The nature of approximations used for derivation of the
quasiclassical formula (\ref{rmat-wid-dms-3s}) is such that they can
not be used for derivation of a corresponding formula for a
``diproton'' model. The formula (\ref{wid-3-bark}) [actually
analogous formula] was used in the papers
\cite{bar01,bro02,bar02,bar03,bro03} in the ``diproton'' form by
formally making the ``X'' subsystem to be subsystem of the $p$-$p$
motion. We find serious problems in such an approach.

We can consider this issue from a different side: the analysis can
be started from the original expressions (\ref{width-3b}),
(\ref{coef-a}) [not from approximation (\ref{rmat-wid-dms-3})] and
the situation studied with one resonant and one nonresonant
subsystem. The direct integration in (\ref{coef-a}) for the case of
one resonant and one nonresonant continuum in the subsystems leads
in the general case to the expression analogous to
(\ref{wid-3-bark}). However it contains a complicated additional
coefficient that is strongly dependent on fine details of the system
geometry and the behaviour of the nonresonant continuum. Thus we can
conclude that in this case Eq.\ (\ref{wid-3-bark}) should be correct
only within an order of the magnitude.

The analysis provided in Ref.\ \cite{gri07} for the case physically
corresponding to the ``diproton'' model (the OFSI model in the ``T''
Jacobi system) showed that the situation in that particular case is
even worse. The semianalytical three-body model of Ref.\
\cite{gri07} provides the width values which are very small
(compared to those typically evaluated in diproton model). They
could be matched to those obtained from  Eq.\ (\ref{wid-3-bark})
only if very small channel radii $r_{chy}$ ($1-2$ fm) are chosen for
emission of the ``diproton''. This requirement is certainly not
consistent with the practice of the ``diproton'' model application,
where the $r_{chy}$ is typically chosen as a nucleus radius plus
some ``radius of diproton''.


\subsection{Relation to the three-body calculations}


Some important points could be understood using the information listed in Table
\ref{tab:wid-sens}.

Treatment of the Coulomb interaction in the present model follows the 
approximations discussed in detail in Ref.\ \cite{gri07}: ``no $p$-$p$'' means 
that Coulomb interaction between protons is just neglected (product of charges 
are $Z_x=Z_y=Z_{core}$ both in $X$ and $Y$ subsystems); effective treatment of 
Coulomb ``Eff.'' means that Coulomb interaction in $Y$ coordinate is formed by 
proton and ``effective particle'' core+proton (product of charges are 
$Z_x=Z_{core}$ in $X$ and $Z_y=Z_{core}+1$ in $Y$ subsystem). Sensitivity to the 
choice of the Coulomb treatment is relatively high (factor $1.5-5$, depending on 
particular model).

The values of $\Gamma$ derived from Eq.\ (\ref{rmat-wid-dms-3s}) reasonably well 
agree with the corresponding results of the three-body calculations with the 
simplified Hamiltonian from Ref.\ \cite{gri07}. The disagreement can be reduced 
if (i) the widths of the subsystems ($\Gamma_x$, $\Gamma_y$) are taken the same 
as in the three-body model, (ii) corrections for coefficient $D_3$ (see Table 
\ref{tab:wid-three-d}) are taken into account and (iii) radii of channels in 
Eq.\ (\ref{rmat-wid-dms-3s}) are close to the inner classical turning points 
\cite{turning} for corresponding potentials in the three-body model (see Table 
\ref{tab:wid-sens}).

The sensitivity of the results to the ``unphysical'' (not leading to
modification of observable values $\Gamma_{xa}$, $\Gamma_{ya}$)
variation of channel radii is moderate (about factor of 1.5) except
for the $\Gamma_{s}$ model Eq.\ (\ref{wid-3-prev}) from our previous
works. However, even this model is providing results consistent with
Eq.\ (\ref{rmat-wid-dms-3s}) if the channel radii are chosen close
to classical turning point.

From comparison of the simplified three-body model with the
realistic three-body model in Ref.\ \cite{gri07} we can see that the
calculations using the effective Coulomb interaction are reasonably
close to realistic results and should be a preferable choice.
However, even calculations with the effective Coulomb interaction
differs from the realistic results typically by a factor $1.3-3$. Thus the
calculations in the quasiclassical model presented here (which
itself is an approximation to the three-body model with simplified
Hamiltonian of Ref.\ \cite{gri07}) are not a replacement for the
realistic three-body calculations if a better than mentioned precision is 
requested.


\section{Conclusion.}


In this work we derived the quasiclassical (``R-matrix type'')
formula for two-proton decay widths. The preexponent coefficients
are defined and evaluated numerically using the simplified
three-body model. The following aspects of the obtained results
should be emphasized.

\noindent(i) The derivation is based on the three-body model with a
simplified Hamiltonian \cite{gri07}, which omits one FSI ($p$-$p$)
and treats another in an approximate way (one of core-$p$
interactions). The first approximation can be justified by the
weakness of the $p$-$p$ interaction compared to core-$p$ interaction
and should be good in heavy systems. The second approximation is
connected with the finite mass of the core and also becomes well
justified in heavy systems.

\noindent(ii) The derivation of the quasiclassical formula requires existence of
\emph{narrow} states in \emph{both} core-$p$ subsystems. This condition is also
well satisfied for the ground states of relatively heavy systems.

\noindent(iii) The derived formula is basically the same as that
obtained by Galitsky and Cheltsov in Ref.\ \cite{gal64} by a
somewhat different procedure. As far as we started from construction
and validation of the simplified three-body model \cite{gri07} we
were able to define a precision of approximations used for
transition to the quasiclassical model and define ingredients of the
model which remained undefined in \cite{gal64}.

\noindent(iv) The most important dependencies of the three-body
width (\ref{rmat-wid-dms-3s}) are fixed by observable properties of
the subsystems (positions and widths of the lowest resonances in the
subsystems). However, there is ``unphysical'' sensitivity to channel
radii when the observables for the subsystems are fixed. The
quasiclassical formula provides a good agreement with the simplified
three-body model when the radii of channels in the subsystems are
chosen to be close and outside of the inner classical turning points of the
barriers.

\noindent(v) The formula (\ref{wid-3-bark}), used in papers
\cite{bar99,bar01,bro02,bar02,bar03,bro03}, is a special case of
Eq.\ (\ref{rmat-wid-dms-3s}). It can be obtained by formally
assuming the energy of one of the states in the subsystems to go to
positive infinity. Such an assumption is unphysical when both valence
protons populate states with the same quantum numbers. Thus formula
(\ref{wid-3-bark}) is valid within a factor which can be as large
as 4.

\noindent(vi) The derived expressions can not be used in the form of
the ``diproton'' model without introducing a large (above an order
of magnitude) uncertainty.

Having in mind the origin and scale of the uncertainties introduced
by reducing the realistic three-body model to a simplified
three-body model (discussed in details in \cite{gri07}) and by
reducing the simplified three-body model to the quasiclassical three-body
model (considered in this work) we now get a basis for the
appropriate (within the limits of its reliability) application of
this model for estimates of the two-proton widths.


\section{Acknowledgements}


We are thankful to Prof. G. Nyman for careful reading of the manuscript and 
useful comments.  The authors acknowledge the financial support from the Royal 
Swedish Academy of Science. LVG is supported INTAS Grants 03-51-4496 and 
05-1000008-8272, Russian RFBR Grants Nos.\ 05-02-16404 and 05-02-17535 and 
Russian Ministry of Industry and Science grant NS-8756.2006.2.



\begin{table*}[h]
\caption{Properties of the test ``$^{2}n+^{2}n$'' system (the
reduced mass is equal to the neutron mass) for square well potential
$R=4$ fm and $E_r=0.1$ MeV. The dimensionless reduced widths
calculated for well boundary ($\theta_{1}^{2}$) and for ``infinity''
in the sense of Eq.\ (\ref{norm-teta-out}) ($\theta_{2}^{2}$).
Values $B_{l}$ [Eq.\ (\ref{en-dep-2})] calculated for different
energies of a resonance in auxiliary Hamiltonian. The width $\Gamma$
obtained by two-body scattering calculations [it exactly coincides
with $\Gamma$ defined by Eq.\ (\ref{r-mat-wid-2}) for reduced widths
$\theta_{2}^{2}$].}
\label{tab:r-mat-2}
\begin{ruledtabular}
\begin{tabular}[c]{cccccccc}
$l$ &  $\theta_{1}^{2}$ & $\theta_{2}^{2}$ & $\theta_{2}^{2}
/\theta_{1}^{2}$ &$B_l(E_a=1.1E_r)$ & $B_l(E_a=2E_r)$ & $B_l(E_a=5E_r)$ &
$\Gamma$ (MeV) \\
\hline
1 & 2.044 & 0.671 & 0.329 & 0.81 & 0.75 & 0.58 & $3.46\times 10^{-2}$\\
2 & 2.013 & 1.205 & 0.598 & 1.04 & 1.06 & 1.11 & $5.58\times 10^{-4}$\\
3 & 2.007 & 1.432 & 0.714 & 1.007& 1.011& 1.03 & $2.14\times 10^{-6}$\\
4 & 2.002 & 1.557 & 0.777 & 1.0022& 1.0046& 1.013 & $3.54\times10^{-9}$
\end{tabular}
\end{ruledtabular}
\end{table*}

\begin{table*}[h]
\caption{Parameters for quasiclassical approximation calculated in a three-body
(TFSI) model Ref.\ \cite{gri07}.}
\label{tab:wid-three-d}
\begin{ruledtabular}
\begin{tabular}[c]{ccccccccccc}
& $E$ & $l_{x}$ & $E_{xa}$ & $l_{y}$ & $E_{ya}$ & $E_{a}$  & $\left\langle
V_{3}\right\rangle ^{2}$ & $\left\langle V_{3}^{2}\right\rangle$  &
$\left\langle
V_{3}^{2}\right\rangle/\left\langle V_{3}\right\rangle ^{2}$  & $D_{3}$\\
\hline
$^{17}$Ne  & 0.344 & 0 & 0.535 & 2 & 0.96 & 1.495 &  1.714 & 2.426 & 1.414 &
1.338 \\
$^{45}$Fe & 1.154 & 1 & 1.480 & 1 & 1.480 & 2.960 &  3.285 & 3.991 & 1.215 &
1.098
\end{tabular}
\end{ruledtabular}
\end{table*}

\begin{table*}[h]
\caption{Width sensitivity to the treatment of the charges in the
subsystems and to channel radii for fixed properties of the
subsystems ``X'' and ``Y''. The three-body widths $\Gamma_{s}$ Eq.\
(\ref{wid-3-prev}), $\Gamma_{b}$ Eq.\ (\ref{wid-3-bark}),
$\Gamma_{m}$ Eq.\ (\ref{gamma-3-m}), $\Gamma$ Eq.\
(\ref{rmat-wid-dms-3s}) with $D_{3}=1$, are given in $10^{-14}$ MeV
units for $^{17}$Ne and $10^{-19}$ MeV units for $^{45}$Fe. The
two-proton decay energies $E_{3r}$ are 0.344 MeV for $^{17}$Ne and
1.154 MeV for $^{45}$Fe. The recent experimental data on $^{45}$Fe
\cite{dos05} provide $\Gamma_{2p}=2.85_{-0.68}^{+0.65}\times
10^{-19}$ MeV [$T_{1/2}(2p) =1.6_{-0.3}^{+0.5}$ ms] for
$E_{3r}=1.154(16)$ MeV and two-proton branching ratio $Br(2p)=0.57$.
The $E_{xa}$, $E_{ya}$ values are given in Table
\ref{tab:wid-three-d}. The widths $\Gamma_{xa}$ and $\Gamma_{ya}$
are chosen to be the same as in the corresponding potential model of
Ref.\ \cite{gri07}, Tables I, II.}
\label{tab:wid-sens}
\begin{ruledtabular}
\begin{tabular}[c]{cccccccccccccc}
& Coulomb & $l_{x}$ & $\theta^2_{x}$ & $r_{chx}$ (fm) & $\Gamma_{xa}$ (keV) &
$l_{y}$ & $\theta^2_{y}$ & $r_{chy}$ (fm) & $\Gamma_{ya}$ (keV) & $\Gamma_{s}$ &
$\Gamma_{b}$ & $\Gamma_{m}$ & $\Gamma$  \\
\hline
$^{17}$Ne & No $p$-$p$ & 0 & 0.986 & 3.53\footnotemark[1] & 17.9 & 2 & 1.85 &
3.53\footnotemark[1] & 3.5 & 2.63 & 1.75 & 3.70 & 3.72 \\
& No $p$-$p$ & 0 & 0.667 & 5.06\footnotemark[2] & 17.9 & 2 & 1.145 &
4.04\footnotemark[2] & 3.5 & 9.44 & 2.04 & 4.33 & 4.35 \\
& Eff.\ & 0 & 0.667 & 5.06\footnotemark[2] & 17.9 & 2 & 1.145 & 4.12 & 2.2 &
1.86 & 0.385 & 0.845 & 0.844 \\
& Eff.\ & 0 & 0.667 & 5.06\footnotemark[2] & 17.9 & 2 & 0.162 &
7.0\footnotemark[3] & 2.2 & 2.27 & 0.510 & 1.12 & 1.10 \\
& Eff.\ & 0 & 0.363 & 8.0\footnotemark[3] & 17.9 & 2 & 1.145 & 4.12  & 2.2 &
5.84 & 0.522 & 1.15 & 1.13 \\
& Eff.\ & 0 & 0.363 & 8.0\footnotemark[3] & 17.9 & 2 & 0.162 &
7.0\footnotemark[3] & 2.2 & 71.2 & 0.691 & 1.52 & 1.50 \\
$^{45}$Fe & No $p$-$p$ & 1 & 1.07 & 4.72\footnotemark[2] & 0.257 & 1 & 1.07 &
4.72\footnotemark[2] & 0.257 & 10.5 & 3.24 & 12.7 & 12.9  \\
& No $p$-$p$ & 1 & 0.888 & 4.94\footnotemark[1] & 0.257 & 1 & 0.888 &
4.94\footnotemark[1] & 0.257 & 16.6 & 3.36 & 13.1 & 13.3 \\
& Eff.\ & 1 & 1.07 & 4.72\footnotemark[2] & 0.257 & 1 & 1.03 & 4.76 & 0.15 &
3.94 & 1.14 & 4.54 & 4.62 \\
& Eff.\ & 1 & 0.888 & 4.94\footnotemark[1] & 0.257 & 1 & 0.85 &
4.98\footnotemark[1] & 0.15 & 6.24 & 1.18 & 4.70 & 4.78 \\
& Eff.\ & 1 & 0.143 & 7.5\footnotemark[3] & 0.257 & 1 & 0.128 & 7.56 & 0.15 &
667 & 2.02 & 8.02 & 8.16
\end{tabular}
\end{ruledtabular}
\footnotetext[1]{This is radius from systematics
$r_{ch}=1.4(A_{\text{core}}+1)^{1/3}$ used in the papers
\cite{gri01a,gri02,gri03,gri03a,gri03c}.}
\footnotetext[2]{This is radius of the classical inner turning point for
potential Ref.\ \cite{gri07}, Tables I, II.}
\footnotetext[3]{At this radius the nuclear potential becomes negligible in
potential model Ref.\ \cite{gri07}, Tables I, II.}
\end{table*}



\begin{thebibliography}{99}
\bibitem{turning}
For $^{17}$Ne $3/2^-$ state the TFSI model width with effective Coulomb is $1.37
\times 10^{-14}$ MeV (Table I of Ref.\ \cite{gri07}) for classical turning
points 5.06 fm in $s$-wave and 4.04 fm in $d$-wave. For $^{45}$Fe $3/2^-$ state
the TFSI model width with effective Coulomb is $4.9 \times 10^{-19}$ MeV (Table
II of Ref.\ \cite{gri07}) for classical turning points around 4.72 fm in both
``X'' and ``Y'' subsystems.
\end{thebibliography}
\end{document}